\documentclass{article}

\def\ParSkip{} 
\usepackage{amssymb,amsmath,amsthm,bbm}
\usepackage{verbatim,float,url,dsfont}
\usepackage{graphicx,subfigure,psfrag}
\usepackage{mathtools,enumitem}
\usepackage{multirow}
\usepackage{ragged2e}
\usepackage{xr-hyper}
\usepackage{array}

\usepackage[colorlinks=true,citecolor=blue,urlcolor=blue,linkcolor=blue]{hyperref}
\usepackage[margin=1in]{geometry}
\usepackage[round]{natbib}

\usepackage[utf8]{inputenc} 
\usepackage[T1]{fontenc}    
\usepackage{booktabs}       
\usepackage{nicefrac}         
\usepackage{microtype}      
\usepackage{amsfonts}
\usepackage{authblk}
\usepackage{algorithm2e}

\ifdefined\TimesFont 
\usepackage{times} 
\fi

\ifdefined\ParSkip 
\usepackage{parskip} 
\fi

\newtheorem{theorem}{Theorem}

\newtheorem{proposition}{Proposition}
\theoremstyle{definition}

\newtheorem*{assumption*}{\assumptionnumber}
\providecommand{\assumptionnumber}{}


\makeatletter
\newcommand*\rel@kern[1]{\kern#1\dimexpr\macc@kerna}
\newcommand*\widebar[1]{%
  \begingroup
  \def\mathaccent##1##2{%
    \rel@kern{0.8}%
    \overline{\rel@kern{-0.8}\macc@nucleus\rel@kern{0.2}}%
    \rel@kern{-0.2}%
  }%
  \macc@depth\@ne
  \let\math@bgroup\@empty \let\math@egroup\macc@set@skewchar
  \mathsurround\z@ \frozen@everymath{\mathgroup\macc@group\relax}%
  \macc@set@skewchar\relax
  \let\mathaccentV\macc@nested@a
  \macc@nested@a\relax111{#1}%
  \endgroup
}
\makeatother

\title{Distribution-free Conformal Prediction for Ordinal Classification}
\author[1]{Subhrasish Chakraborty}
\author[1]{Chhavi Tyagi}
\author[2]{Haiyan Qiao}
\author[1]{Wenge Guo\thanks{Author e-mail addresses: sc2325@njit.edu, ct364@njit.edu, hqiao@csusb.edu, wenge.guo@njit.edu}}
\affil[1]{Department of Mathematical Sciences\\
       New Jersey Institute of Technology}
\affil[2]{School of Computer Science \& Engineering\\
       California State University San Bernardino}

\date{\today}

\begin{document}
\maketitle

\begin{abstract}
Conformal prediction is a general distribution-free approach for constructing prediction sets combined with any machine learning algorithm that achieve valid marginal or conditional coverage in finite samples. Ordinal classification is common in real applications where the target variable has natural ordering among the class labels. In this paper, we discuss constructing distribution-free prediction sets for such ordinal classification problems by leveraging the ideas of conformal prediction and multiple testing with FWER control.  
Newer conformal prediction methods are developed for constructing contiguous and non-contiguous prediction sets based on marginal and conditional (class-specific) conformal $p$-values, respectively. Theoretically, we prove that the proposed methods respectively achieve satisfactory levels of marginal and class-specific conditional coverages. Through simulation study and real data analysis, these proposed methods show promising performance compared to the existing conformal method. 
\end{abstract}


\section{Introduction}
\label{sec:intro}

Ordinal classification, also known as ordinal regression or ordinal prediction, is a machine learning task that involves predicting a target variable with ordered categories \citep{mccullagh1980, agresti2010analysis}. In ordinal classification, the target variable has a natural ordering or hierarchy among its categories, but the intervals between the categories may not be evenly spaced or defined. Unlike regular classification, where the classes are nominal and unordered, ordinal classification takes into account the ordering relationship between the classes. This makes it suitable for situations where the outcome variable has multiple levels of severity, satisfaction ratings, or rankings. Here are a few examples of ordinal classification: customer satisfaction levels, movie ratings, disease severity levels, and education levels \citep{agresti2010analysis}. 

In ordinal classification, the goal is to learn a model that can accurately predict the ordinal variable's value given a set of input features. The model needs to understand the ordering of the classes and make predictions that respect this order. In the literature, 
some conventional classification algorithms have been adapted or modified to address ordinal classification, for example, ordinal logistic regression, SVM, decision trees, random forest, and neural networks \citep{harrell2015ordinal, da2010all, kramer2001prediction, janitza2016random, cheng2008neural}. Some alternative methods are also specifically developed for ordinal classification problems by fully exploiting the ordinal structure of the response variables
\citep{frank2001simple, cardoso2007learning, gutierrez2015ordinal}. However, these existing methods can only provide point prediction, which is not adequate in some high
stakes areas such as medical diagnosis and automatic driving. Uncertainty quantification (UQ) techniques aim to go beyond point predictions and provide additional information about the reliability of these predictions. There are various techniques for UQ in machine learning, including Bayesian methods, calibration, and conformal prediction \citep{hüllermeier2021}. 

Conformal prediction is a unique distribution-free UQ technique that provides a prediction set rather than a point prediction for the true response with guaranteed coverage \citep{Vovk1999, vovk2005, shafer2008, angelopoulos2021, fontana2023}. It can be used as a wrapper with any underlying algorithm. In this paper, we use the conformal prediction technique to construct prediction sets for 
ordinal classification problems. By combining the ideas of conformal prediction and multiple testing, two new conformal prediction methods are introduced for constructing contiguous and non-contiguous prediction sets.
Firstly, the problem of ordinal classification is reformulated as a problem of multiple testing; Secondly, for each constructed hypothesis, the marginal and conditional conformal $p$-values are respectively calculated; Thirdly, based on these marginal (conditional) conformal $p$-values, three multiple testing procedures are developed for controlling marginal (conditional) familywise error rate (FWER); Finally, based on the testing outcomes of these procedures, the prediction sets are constructed and proved having guaranteed marginal (conditional) coverage. There are almost no works of applying conformal prediction to address ordinal classification in the literature. To our knowledge, \cite{lu2022} is the only existing work, in which, a new (split) conformal prediction method is developed for constructing adaptive contiguous prediction region. This method is proved to have guaranteed marginal coverage, however, it cannot guarantee to have more desired conditional coverage. Moreover, it does not work well for high dimensional data. 
Compared to the method introduced in \cite{lu2022}, our proposed methods generally show via theoretical and numerical studies
better performance in the settings of higher dimensions and in terms of class-specific conditional coverage; especially for the conditional conformal $p$-values based methods, they are proved to have guaranteed conditional coverage. 

The rest of this paper is structured as follows. In Section 2, we briefly introduce split-conformal prediction and review related works, followed by Section 3 which presents the development
of our proposed conformal methods using the idea of multiple testing. Section 4 provides numerical studies to evaluate the performance of the proposed methods compared to the existing method. Some discussions are presented in Section 5 and all proofs are deferred to the appendix section.

\section{Preliminaries}\label{sec:cp}

In this section, we briefly describe the conformal prediction framework and review the related literature.
\subsection{Conformal Prediction}\label{cp}

Conformal prediction is a general statistical technique to construct prediction sets for any supervised learning method. The main advantage of this approach is that it is distribution-free and can work with any underlying algorithm. Conformal prediction is broadly of two types -- full conformal prediction and split-conformal prediction. The full conformal prediction uses all the observations to train the underlying algorithms \citep{vovk2005}. In contrast, split-conformal prediction \citep{papadopoulos2002} involves splitting the training data into proper training data to train the underlying algorithm and calibration data to calculate the threshold for forming prediction sets. Our proposed methods are based on the split-conformal method.
Consider a multi-class classification problem with feature space $ \mathcal{X} $ and labels $ \mathcal{Y} = \{ 1,2,\cdots,K \} $. Given the training observations  $ (X_i,Y_i)_{i=1}^{2n}$ and a test input $X_{2n+1}$, the goal is to find a prediction set $ C(X_{2n+1}) : \mathcal{X} \mapsto  2^{\mathcal{Y}} $ that contains the unknown response $Y_{2n+1} $ with enough statistical coverage.

The split-conformal procedure suggests to split $ 2n $ observations to $ n $ training observations, i.e., $ (X_i,Y_i)_{i=1}^{n} $, which are used to train $ \hat{f}$, an underlying classifier such that $ \hat{f}: \mathcal{X} \mapsto \mathcal{Y} $ and the remaining $ n $ observations $ (X_i,Y_i)_{i=n+1}^{2n} $ for calibration. The central part of this technique involves calculating the conformity scores for each observation, which measures how much the test observation conforms with the calibration observations. There can be several choices of conformity scores for multi-class classification problem, including posterior class probability, cumulative probability, and regularized cumulative probability \citep{sadinle2019, romano2020, Angelopoulos2020}. Given the score function $ s : \mathcal{X} \times \mathcal{Y} \mapsto \mathbb{R} $, the conformity score for the $ i^{th} $ calibration observation is defined as $s_i = s(X_i,Y_i), \hspace{0.2cm} i=n+1,\cdots,2n$.

For a test input $ X_{2n+1}, $ we compute the conformity score for each class label. Therefore, for a class label $ y \in \mathcal{Y} $ the conformity score corresponding to $ (X_{2n+1},y) $ is $ s_{2n+1} = s(X_{2n+1},y) $. By using the conformity scores obtained for the calibration observations and the test input coupled with a given label $y$,  we can calculate the conformal $p$-value to test whether the unknown true label $Y_{2n+1}$ corresponding to the test input $X_{2n+1}$ is $y$ or not. The (marginal) conformal $p$-value is defined as,
\begin{equation}\label{eq:3}
    p(X_{2n+1}, y) = \frac{\sum\limits_{i=n+1}^{2n} \mathbb{I}\left(s_{2n+1} \leq s_i\right) + 1}{n+1}. 
\end{equation}
The final step involves constructing the prediction set $ C(X_{2n+1}) = \{y:p(X_{2n+1}, y) \geq \alpha \} $, which satisfies
\begin{equation}\label{eq:4}
 \mathbb{P}(Y_{2n+1} \in C(X_{2n+1})) \geq 1 - \alpha,
\end{equation}
when the calibration and test observations $ (X_i,Y_i)_{i=n+1}^{2n+1}$ are exchangeably distributed, where $ \alpha \in (0,1)$ is a pre-specified mis-coverage level. Equation (\ref{eq:4}) is called \textit{marginal validity} of the prediction set $C(X_{2n+1})$. It guarantees that the true label $Y_{2n+1}$ is contained in the prediction set with $ 100(1 - \alpha)\% $ confidence. \cite{vovk2012conditional} introduced another type of conformal $p$-value which is called as the conditional conformal $p$-value. Let $\mathcal{D}_{cal} = \{n+1, \ldots, 2n\}$ denote the indices of the calibration observations 
$ (X_i,Y_i)_{i=n+1}^{2n} $. For a test input $ X_{2n+1} $ and any class $y=1,\dots,K$, the (class-specific) conditional conformal $p$-value given $Y_{2n+1} = y$ is defined as 
\begin{equation}\label{eq:conditional pval}
    p(X_{2n+1}|y) = \frac{\sum_{i\in \mathcal{I}_y}\mathbb{I} \{s_{i} \leq s_{2n+1}\} + 1}{n_y + 1}, 
\end{equation}
where $\mathcal{I}_{y} = \{ i \in \mathcal{D}_{cal}: Y_i = y \}$, $n_y = |\mathcal{I}_{y}|$ is the size of $\mathcal{I}_{y}$, and
$s_{i} = s(X_i,y)$ for $i=n+1, \ldots, 2n+1$.

In general, the concept of conditional coverage such as \textit{object conditional validity} and \textit{class-specific conditional validity} are more relevant to practical applications \citep{vovk2012conditional, lei2014, foygel2021limits}. If conditional coverage is guaranteed, our predictions are valid for a more specific
sub-population, not merely on average over a general population. Specifically, in classification problems, \textit{class-specific conditional validity} provides conditional coverage for each given class, which is defined as
\begin{equation}\label{eq:5}
    \mathbb{P}(Y_{2n+1} \in C(X_{2n+1})|Y_{2n+1} = y) \geq 1 - \alpha
\end{equation}
for any $ y=1,\dots,K$.
Proposition 1 and 2 below ensure that the marginal and conditional conformal $p$-values are valid, which result in desired marginal and (class-specific) conditional coverage.

To simplify the notation, we let $Z_i = (X_i,Y_i)$ for $i=1, \ldots, 2n+1$ and denote the conformity scores of the calibration data, $\{Z_i\}_{i=n+1}^{2n}$, as $s_i$'s, and the conformity score of the test data, $Z_{2n+1}=(X_{2n+1},Y_{2n+1})$, as $s_{2n+1}$, where $Y_{2n+1}$ is unknown. These notations are used in all propositions and theorems presented in this paper.

\begin{proposition}
Suppose that $ \{Z_i\}_{i=1}^{2n+1} $ where $ Z_i = (X_i,Y_i) $ are exchangeable random variables, then the marginal conformal $p$-values defined below as,

\begin{equation}\label{eq:p_val}
    p(Z_{2n+1}) = \frac{ \sum_{i=n+1}^{2n} \mathbb{I}(s_i \leq s_{2n+1}) + 1}{n+1}
\end{equation}  
is valid in the sense that for any $ t \in [0,1],  $ we have  
$$ \mathbb{P}(p(Z_{2n+1}) \leq t) \leq t. $$
Moreover, if the conformity scores $ \{s_i\}_{i=n+1}^{2n+1} $ are distinct surely, we have, $$ p(Z_{2n+1}) \sim U\Bigl\{\frac{1}{n+1},\cdots,1\Bigl\} $$
\end{proposition}
\label{prop_1}

\begin{proposition}
Suppose that $ \{Z_i\}_{i=1}^{2n+1} $ where $ Z_i = (X_i,Y_i) $ are exchangeable random variables, then for any $ y \in \mathcal{Y}  $, given $ \mathcal{I}_y \subseteq \mathcal{D}_{cal}$ and $ Y_{2n+1} = y $, the corresponding conditional conformal $ p$-value as defined in equation (\ref{eq:conditional pval}), is conditionally valid in the sense that for any $ t \in [0,1] $, 
$$ \mathbb{P}\left(p(X_{2n+1}|y) \leq t ~ \big|~ \mathcal{I}_y,Y_{2n+1} = y\right) \leq t. $$
Moreover, if $ \{s_{i}\}_{i\in \mathcal{I}_y \cup \{2n+1\}} $ are distinct surely, we have that conditional on $\mathcal{I}_y$ and $ Y_{2n+1} = y $,
$$ p(X_{2n+1}|y) \sim U\Bigl\{ \frac{1}{n_y+1},\cdots,1\Bigl\}.$$
\end{proposition}
\label{prop_2}

\subsection{Related work}\label{sec:2.2}
The framework of Conformal prediction was introduced by Vladimir Vovk and his collaborators \cite{Vovk1999, vovk2005} and has found many applications in classification problems. \cite{shafer2008} and \cite{angelopoulos2021} provided a tutorial introduction and  brief literature review on this field. Several conformal methods have been developed to address binary classification \citep{lei2014} and multi-class classification problems \citep{hechtlinger2018cautious, sadinle2019, romano2020, Angelopoulos2020, tyagi2023}. Coverage guarantees of all these methods are established under the assumption of exchangeability. Very recently, some new conformal prediction methods have been developed in the settings of non-exchangeability  \citep{tibshirani2019conformal, gibbs2021adaptive, barber2023conformal}. 

Although various conformal prediction methods have been developed for conventional classification problems, however, to our knowledge, \cite{lu2022} is the only reference that is specifically devoted to address ordinal classification problems using conformal prediction methods, in which an adaptive conformal method is developed for constructing contiguous prediction sets for ordinal response and is applied to AI disease rating in medical imaging. In addition, \cite{xu2023} is the closely related reference in which newer methods are developed for two types of loss functions specially designed for ordinal classification
in the more general framework of conformal risk control.

\section{Method}\label{sec:TB-MLCP}

In  this section, we introduce several new conformal prediction methods for ordinal classification problems, in which there is a natural ordering among the classes labels. For simplicity, we assume a descending order of priority from class $1$ to $K$ in the response space $ \mathcal{Y} = \{ 1,2,\cdots,K \} $.

\subsection{Problem Formulation}

We formulate the ordinal classification problem as a multiple testing problem. Specifically, by using the One-vs-All (OVA) strategy \citep{Rifkin2004},  for each class label, we construct a hypothesis to test whether or not a given test input $X_{2n+1}$ belongs to the particular class. The construction of the hypothesis is described as follows,
\begin{equation}\label{eq:hyp}
\begin{centering}
    H_i : Y_{2n+1} = i  \hspace{4mm} \textnormal{ vs } \hspace{4mm} H'_i : Y_{2n+1} \neq i, 
\end{centering}
\end{equation}
for $i=1,\cdots,K.$ 
It is easy to see that all these hypotheses are random and there is only one true null. To test each individual hypothesis $H_i$, we use the corresponding marginal conformal $p$-value $p(X_{2n+1}, i)$; to test $H_1, \ldots, H_K$ simultaneously, we consider the following three $p$-value-based testing procedures:
\begin{itemize}
    \item \textbf{Procedure 1} : Test $ H_1, H_2, \cdots, H_K $ sequentially. The test is performed as follows.
    \begin{itemize}
        \item If $p(X_{2n+1}, 1) \leq \alpha $, reject $ H_1 $, move to test  $ H_2 $ else stop testing;
        \item For $ i = 2,\cdots, K-1 $, if $ p(X_{2n+1}, i) \leq \alpha $, reject $ H_i $, move to test $ H_{i+1}$ else stop testing;
        \item If $p(X_{2n+1}, K) \leq \alpha $, reject $ H_K $ else stop testing.
    \end{itemize}
    \item \textbf{Procedure 2} : Test $ H_K, H_{K-1},\cdots, H_1 $ sequentially. The test is performed as follows.
    \begin{itemize}
        \item If $p(X_{2n+1}, K) \leq \alpha $, reject $ H_K $, move to test $ H_{K-1} $ else stop testing;
        \item For $ i = K-1,\cdots, 2 $, if $p(X_{2n+1}, i) \leq \alpha $, reject $ H_i $, move to test $ H_{i-1}$ else stop testing;
        \item If $p(X_{2n+1}, 1) \leq \alpha $, reject $ H_1 $ else stop testing. 
    \end{itemize}
    \item \textbf{Procedure 3} : Single-step procedure with common critical value $\alpha $. This procedure rejects any hypothesis $H_i$ if and only if $p(X_{2n+1}, i) \leq \alpha $.
\end{itemize}

Procedure 1 and 2 are two pre-ordered testing procedures for which Procedure 1 follows the same testing order as that of the $K$ classes  whereas 
Procedure 2 uses the reverse order of these classes \citep{dmitrienko2009multiple}.  Procedure 3 is actually a conventional Bonferroni procedure for a single true null. 
Since there is only one true null among the $K$ tested hypotheses, by Proposition 1, we have that all these three (marginal) conformal $p$-value based procedures strongly control family-wise error rate (FWER) at a pre-specified level $\alpha$ \citep{dmitrienko2009multiple}. For each Procedure $i, i = 1, 2, 3$ defined above, the index set $ A_i $ of the accepted hypotheses is described as follows,
\begin{enumerate}
    \item $ A_1 = \{y_{min}, y_{min}+1,\cdots, K \} $, where $ y_{min} = \min\{y\in \mathcal{Y}:p(X_{2n+1}, y) > \alpha \} $;
    \item $ A_2 = \{ 1, 2, \cdots, y_{max}\} $, where $ y_{max} = \max\{y\in \mathcal{Y}:p(X_{2n+1}, y) > \alpha\} $;
    \item $ A_3 = \{y: p(X_{2n+1}, y) > \alpha \} $.
\end{enumerate}

\subsection{Ordinal Prediction Interval}
\begin{centering}
\begin{figure}
    \centering
    \includegraphics[scale = 0.85]{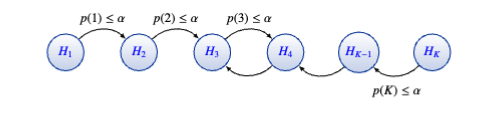}
\caption{Graphical representation of Ordinal Prediction Interval (OPI) with $ K $ nulls where $p(i)$ represents the conformal $p$-values.}
\label{fig:testing_procedure}

\end{figure}
\end{centering}

Based on the the acceptance sets $A_1$ and $A_2$ of Procedure 1 and 2 given as above, we can obtain a new acceptance region $A_{12} = A_1 \cap A_2 = \{y_{min}, \ldots, y_{max}\}$, which is used to define the prediction region $C(X_{2n+1})$ for the unknown response $Y_{2n+1}$. Specifically, the prediction region  $C(X_{2n+1})$ consists of the class labels for which the corresponding hypotheses are both accepted by Procedure 1 and 2, resulting in a \textit{contiguous set} of labels $\{y_{min}, \ldots, y_{max}\}$. This prediction region is referred to as a \textit{prediction interval} in this context. The procedure for constructing the prediction interval $C(X_{2n+1})$ is summarized in Algorithm \ref{alg:one} and illustrated in Figure \ref{fig:testing_procedure}. 

\RestyleAlgo{ruled}
\SetKwComment{Comment}{/* }{ */}

\begin{algorithm}
\caption{Ordinal Prediction Interval}
\label{alg:one}

\KwIn{training set $ \mathcal{D}_{train}=(X_i,Y_i)_{i=1}^n $, calibration set $ \mathcal{D}_{cal} = (X_i,Y_i)_{i=n+1}^{2n}$, test input $ X_{2n+1}$,  black-box algorithm $\mathcal{A}$, conformity score function $s$, mis-coverage level $\alpha.$}
 \KwOut{Prediction interval, $C(X_{2n+1}).$}
Train a classifier $ \mathcal{A}$ on $ \mathcal{D}_{train}$\;
   \For{$ (X_i,Y_i) \in \mathcal{D}_{cal}$}
     {Compute conformity score $ s_{i} = s(X_i,Y_i)$;}
    For each $y \in \mathcal{Y}$, compute conformity score $ s_{2n+1} = s(X_{2n+1},y)$ and corresponding conformal $p$-value $ p(X_{2n+1}, y) $ using equation (1)\; 
$ y_{min} = \min\{ y \in \mathcal{Y} : p(X_{2n+1}, y) > \alpha \} $ \; 
$ y_{max} = \max\{ y \in \mathcal{Y} : p(X_{2n+1}, y) > \alpha \} $ \;
Prediction interval, $C(X_{2n+1}) = \{y_{min},\ldots,y_{max}\}$.\}
\end{algorithm}

\subsection{Ordinal Prediction Set}
Our second method for constructing ordinal prediction regions is based on Procedure 3. In this method, the prediction region $C(X_{2n+1})$ is defined simply using the acceptance region $A_3$ of Procedure 3, that is, $C(X_{2n+1}) = \{y \in \mathcal{Y}: p(X_{2n+1}, y) > \alpha\}$. Specifically, the prediction region consists of any class labels for which the corresponding hypotheses are not rejected by Procedure 3, resulting in a \textit{non-contiguous set} of labels. 
This prediction region is referred to as a \textit{prediction set} in this context. The procedure for constructing the prediction set is detailed in Algorithm \ref{alg:two} below.

\RestyleAlgo{ruled}
\SetKwComment{Comment}{/* }{ */}

\begin{algorithm}
\caption{Ordinal Prediction Set}
\label{alg:two}
 \KwIn{training set $ \mathcal{D}_{train}=(X_i,Y_i)_{i=1}^n $, calibration set $ \mathcal{D}_{cal} = (X_i,Y_i)_{i=n+1}^{2n}$, test observation $ X_{2n+1}$,  black-box algorithm $\mathcal{A}$, conformity score function $s$, mis-coverage level $\alpha.$}
\KwOut{Prediction set, $C(X_{2n+1}).$}
Train a classifier $ \mathcal{A}$ on $ \mathcal{D}_{train}$\;
   \For{$ (X_i,Y_i) \in \mathcal{D}_{cal}$}
      {Compute conformity score $ s_{i} = s(X_i,Y_i)$;}
    For each $y \in \mathcal{Y}$, compute conformity score $ s_{2n+1} = s(X_{2n+1},y)$ and corresponding conformal $p$-value $ p(X_{2n+1}, y) $ using equation (1)\;
Prediction Set, $ C(X_{2n+1}) = \{ y: p(X_{2n+1}, y) > \alpha \}$. \
\end{algorithm}

\noindent In the following, we present two results regarding the FWER control of Procedure 1-3 and (marginal) coverage guarantees of Algorithm 1-2 introduced as above. 

\begin{proposition}
Suppose that $(X_i,Y_i)_{i=n+1}^{2n+1}$ are exchangeable random variables, then Procedure 1-3 based on marginal conformal $ p$-values, all strongly control the FWER at level $ \alpha $, i.e.,  $FWER \leq \alpha $.
Specifically, if the conformity scores $ \{s_i\}_{i=n+1}^{2n+1} $ are distinct surely, then for Procedure 3, we also have,
$FWER \geq \alpha - \frac{1}{n+1}.$
\end{proposition}

\begin{theorem}
Suppose that $(X_i,Y_i)_{i=n+1}^{2n+1}$ are exchangeable random variables, then the prediction region $ C(X_{2n+1}) $ determined by Algorithm 1 and 2 both satisfy
$$ \mathbb{P}(Y_{2n+1} \in C(X_{2n+1})) \geq 1 - \alpha.$$
Specifically, for $C(X_{2n+1}) $ determined by Algorithm 2, if the conformity scores $ \{s_i\}_{i=n+1}^{2n+1} $ are distinct surely, we have 
$$ \mathbb{P}(Y_{2n+1} \in C(X_{2n+1})) \leq 1 - \alpha + \frac{1}{n+1}.$$
\end{theorem}

\subsection{Class-specific conditional coverage}

 To achieve more desired class-specific conditional coverage for our constructed prediction intervals and prediction sets, in Procedure 1-3 we use (class-specific) conditional conformal $p$-values $p(X_{2n+1}|y)$ as described in equation (\ref{eq:conditional pval}), instead of the marginal conformal $p$-values $p(X_{2n+1}, y)$ for simultaneously testing $H_1, \ldots, H_K$ formulated in equation (\ref{eq:hyp}). It is shown in Proposition 4 below that all these three modified procedures strongly control the conditional familywise error rate (FWER) at level $\alpha$, i.e., $\text{FWER}_y = \mathbb{P}(V > 0|Y_{2n+1} = y) \le \alpha$, for any $y\in\mathcal{Y}$, where $V$ is the number of type 1 errors. This result in turn leads to that the prediction regions $C(X_{n+1}|y)$ constructed by Algorithm 1 and 2 based on $p(X_{2n+1}|y)$ satisfy more desired (class-specific) conditional coverage, as stated in Theorem 2. 

\begin{proposition}
Under the same exchangeability assumption as in Proposition 2, Procedure 1-3 based on conditional conformal $p$-values $p(X_{2n+1}|y)$ all strongly control the conditional FWER at level $ \alpha $, i.e., for any $ y \in \mathcal{Y} $, 
$$ FWER_y = \mathbb{P}\{\textnormal{reject}~H_y|Y_{2n+1} = y\} \leq \alpha. $$
Specifically, if the conformity scores $ \{s_{i}\}_{i\in \mathcal{I}_y \cup \{2n+1\}} $ are distinct surely, then for Procedure 3 based on $p(X_{2n+1}|y)$, we have that for any $ y \in \mathcal{Y}$ and $ \mathcal{I}_y \subseteq \mathcal{D}_{cal}$.
$$ \mathbb{P}(\textnormal{reject}~H_y|Y_{2n+1}=y,\mathcal{I}_y ) \geq \alpha - \frac{1}{n_y + 1}.$$ 
\end{proposition}

\begin{theorem}
Under the same exchangeability assumption as in Theorem 1, the prediction region $ C(X_{2n+1}|y)$ determined by Algorithm 1 or 2 based on conditional conformal $p$-values $p(X_{2n+1}|y)$ satisfies
$$ \mathbb{P}\left(Y_{2n+1} \in C(X_{2n+1}|y)~\big |Y_{2n+1} = y\right)~\geq 1 - \alpha $$ for any $ y \in \mathcal{Y}$.
Specifically, for the prediction set $ C(X_{n+1}|y) $ determined by Algorithm 2 based on $p(X_{2n+1}|y)$, if the conformity scores $ \{s_{i}\}_{i\in \mathcal{I}_y \cup \{2n+1\}} $ are distinct surely, we have
$$ \mathbb{P}\left(Y_{2n+1} \in C(X_{2n+1}|y)~\big | Y_{n+1}=y, \mathcal{I}_y \right) \leq 1 - \alpha + \frac{1}{n_y+1}$$ 
for any $ y \in \mathcal{Y}$ and $ \mathcal{I}_y \subseteq \mathcal{D}_{cal}$.
\end{theorem}

\section{Numerical Study}

\noindent In this section, we evaluate the performance of our four proposed methods, Ordinal Prediction Interval (OPI) in Algorithm 1 based on marginal conformal $p$-values (marginal OPI), the OPI based on conditional conformal $p$-values (conditional OPI), Ordinal Prediction Set (OPS) in Algorithm 2 based on marginal conformal $p$-values (marginal OPS), and the OPS based on conditional conformal $p$-values (conditional OPS), in comparison with the existing counterpart developed in \cite{lu2022}, Ordinal Adaptive Prediction Set (OAPS), on simulated data and one real dataset. The comparison is based on the marginal coverage, average set size, and class-specific conditional coverage of the prediction regions for a pre-specified level $\alpha$. The empirical metric we use to measure the class-specific conditional coverage (CCV) of the above methods is defined as
$$CCV = \max\limits_{y \in \{ 1,2,\cdots,K\}} \{ (1-\alpha) - P_y,0\},$$
where $ P_y $ is the estimate of $ \mathbb{P}(Y \in C(X_{2n+1}|y)~|Y = y) $  and $ C(X_{2n+1}|y)$ is the prediction region obtained from Algorithm 1 or 2 using conditional conformal $p$-values for any $ y \in \mathcal{Y} $. Intuitively, the metric measures the maximum of the deviance of the conditional coverage for each of the classes from the desired level of conditional coverage $ 1 - \alpha $. 

In the whole numerical investigations including simulation studies and real data analysis, we use the logistic regression algorithm as the black-box algorithm for our experiments and compute the conformity scores as estimated posterior probabilities of classes. 

\subsection{Simulations}

\noindent We present the simulation study to evaluate the performance of our proposed methods along with the existing method. We consider two simulation settings below, a Gaussian mixture model and a sparse model. 

\begin{enumerate}
    
    \item \textbf{Gaussian mixture.} $(X|Y=k) \sim \pi_1 N_d(\mu_k,\Sigma) + \pi_2 N_d(\mu_{k+1},\Sigma)$ for $ k = 1,2,3 $ and $ (X|Y = 4) \sim \pi_1 N(\mu_4;\Sigma) + \pi_2 N(\mu_1;\Sigma) $ with $\pi_1 = 0.2, \pi_2 = 0.8 $.
\end{enumerate}   
    
\noindent In the above setting, we set $d=2$, $ \mu_1 = (-1,0)$, $ \mu_2 = (-1,-1) $, $ \mu_3 = (0,-1) $, $ \mu_4 = (1,-1)$,  and $ \Sigma $ as the equal correlation matrix with correlation $ = 0.1 $.

2. \textbf{Sparse model}. The sparse model is generated with different dimensions of feature vector with $d = 5, 10, 20, 50, \text{ or } 100$. The features are generated with $ X_i \sim N(0,1) $ with $ Cov(X_i,X_j) = 0.5 $ for $ i \neq j.$ The class labels are generated using the sigmoid function $f(x)$ and the the following decision rule,
    $$ Y_i = k \hspace{4mm} if  \hspace{4mm}  \frac{k-1}{4} \leq f(x_i) < \frac{k}{4} , \hspace{2mm}  where \hspace{2mm} k = 1,2,3,4,$$
where $f(x) = {1}/{(1 + e^{- \beta^{'} x})}$ with $\beta = (\beta_1,\cdots,\beta_d)^{\prime}$ and $x = (x_1, \ldots, x_d)^{\prime}$.  
The value of $\beta$ is set as $(\beta_1,\ldots,\beta_5) = (1,1,1,-\sqrt{2},1)$, and $\beta_i = 0$ for any $5 < i \le d$. 

The sample size for these two simulation settings is 2,000, out of which 500 samples have been used to train the classifier, 525 observations for calibration, and 975 for validation. The simulations are repeated 500 times, and the results are averaged to obtain the final performance metrics.

\begin{figure}
    \centering
    \includegraphics[width = 12cm]{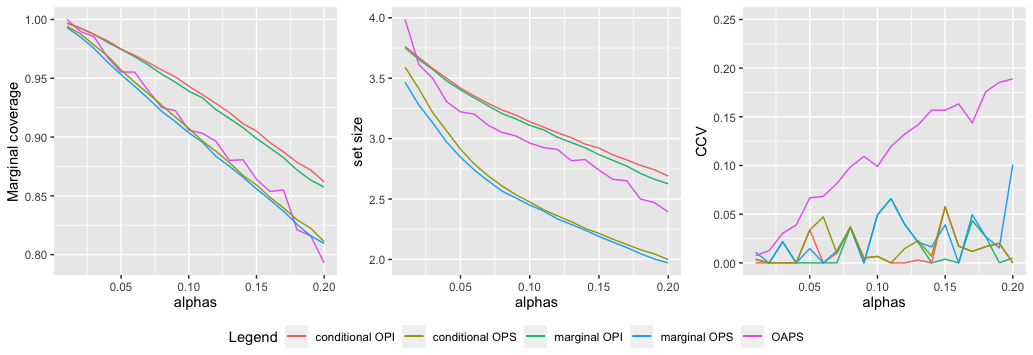}
    \caption{Performance comparison of four proposed methods OPS and OPI using marginal and conditional conformal $p$-values with the existing Ordinal APS under simulation setting 1 in terms of marginal coverage (left), average set size (middle) and class-specific conditional coverage (right) of the prediction sets.}
    \label{fig:gaussian_mixture}
\end{figure}
Figure \ref{fig:gaussian_mixture} displays the performance of our proposed methods along with the existing method under simulation setting 1. It can be seen from the left panel of Figure \ref{fig:gaussian_mixture} that all these five methods empirically achieve the desired level of marginal coverage. The middle panel of Figure \ref{fig:gaussian_mixture} compares the set sizes of the prediction regions corresponding to these five methods. It can be seen from the figure that the marginal OPS and the conditional OPS use shorter set sizes to attain the proper marginal coverage than the existing OAPS  whereas the  OAPS has shorter set sizes than the marginal OPI and the conditional OPI.  Finally, while the marginal coverage is guaranteed by all the methods, the right panel of Figure \ref{fig:gaussian_mixture} shows their differences in class-specific conditional coverage; 
the existing OAPS exhibits the largest value of CCV compared to the proposed methods and among the four proposed methods, the conditional OPI and conditional OPS exhibit lower values of CCV than the marginal OPI and marginal OPS.

\begin{figure}
    \centering
    \includegraphics[width=12cm]{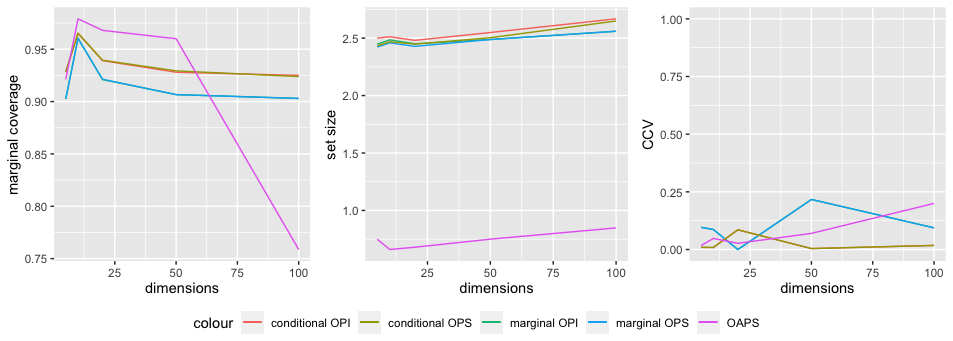}
    \caption{Performance comparison of four proposed methods with existing Ordinal APS under simulation setting 2 with different dimensions of inputs and fixed mis-coverage level $\alpha = 0.1$ 
    in terms of marginal coverage (left), average set size (middle) and CCV (right) of the prediction sets.}
    \label{fig:dimension_versus_performance}
\end{figure}

Figure \ref{fig:dimension_versus_performance} shows the performance of our proposed methods along with the existing method  under simulation setting 2 with different dimensions of inputs. It can be seen from the left panel of this figure that all these methods empirically achieve desired marginal coverage for lower dimensions, however, the existing OAPS massively undercovers for higher dimensions and thus loses the control of mis-coverage rate. From the middle and right panels of Figure \ref{fig:dimension_versus_performance}, we can also see that the conditional OPI and conditional OPS achieve lower values of CCV than the existing OAPS, although the OAPS has the lower set sizes than our proposed methods
for various dimensions of input. 

\subsection{Application to real data}

\begin{figure}
    \centering
    \includegraphics[width=12cm]{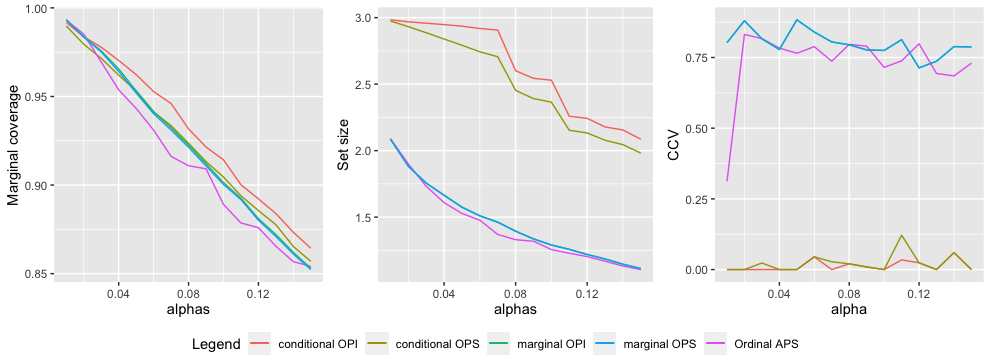}
    \caption{Performance comparison of four proposed methods OPS and OPI using marginal and conditional conformal $p$-values with the existing Ordinal APS in terms of marginal coverage (left), average set size (middle) and CCV (right) of the prediction regions for Traffic Accident Data.}
    \label{fig:traffic_data}
\end{figure}

We also evaluate the performance of our proposed methods on a real dataset, Traffic accident data, which is publicly available on the website of Leeds City Council\footnote{The traffic accident dataset is taken from Leeds Council and its web link url is \url{https://www.leeds.gov.uk/parking-roads-and-travel/connecting-leeds-and-transforming-travel/road-safety/road-traffic-collision-statistics}.}. The real data consists of 1,908 traffic accidents that occurred in the year 2019. The objective is to predict the severity of the casualties, which are classified into three categories -- mild, serious, and fatal based on the features available. In the numerical experiment, 500 observations are used to train the logistic regression model, $ 35\% $ of the remaining observations are used for calibration, and $ 65\% $ for validation. Figure \ref{fig:traffic_data} shows that all these methods empirically achieve desired marginal coverage for different levels of mis-coverage, however, the proposed marginal OPI and marginal OPS, and existing OAPS have lower set sizes than 
the conditional OPI and conditional OPS. It is also evident from Figure \ref{fig:traffic_data} that the proposed conditional OPS and conditional OPI both attain desired class-specific conditional coverage unlike the existing method, OAPS, which seems to largely deviate from the desired level of conditional coverage. 

\section{Concluding Remarks}

In this paper, we discussed the ordinal classification problem in the framework of conformal prediction and introduced two types of conformal methods, OPI and OPS, for constructing distribution-free \textit{contiguous} prediction regions and \textit{non-contiguous} prediction sets, respectively. These methods are developed by leveraging the idea of 
multiple testing with the FWER control and are specifically designed based on marginal conformal $p$-values and (class-specific) conditional conformal $p$-values, respectively. Theoretically, it was proved that the proposed methods based on marginal and conditional $p$-values respectively achieve satisfactory levels of marginal and class-specific conditional coverages. Through some numerical investigations including simulations and real data analysis, our proposed methods show promising results for the settings of higher dimensions and for class-specific conditional coverage. 

This paper discussed constructing valid prediction set for single test input. It would be interesting to discuss how to construct (simultaneous) prediction sets for multiple test inputs with some overall error control such as false discovery rate (FDR) control. 
Another interesting extension might be to relax the conventional distributional assumption we used for classification problems, for which the training data and the test data follow from the same distribution. It will be interesting to see whether the proposed methods can be extended to the settings of distribution shift where the training and test data sets have different distributions. 

\bibliographystyle{plainnat}
\bibliography{main}

\newpage
\appendix

\section{Proofs}

\subsection{Proof of Proposition 1}

\begin{proof}
Suppose, for any given values of conformity scores, $ v_1, \cdots, v_{n+1}, $  they can be rearranged as 
$ \tilde{v}_1 < \cdots < \tilde{v}_k $ 
with repetitions $ n_i $ of $ \tilde{v}_i $ such that $ \sum_{i=1}^{k} n_i = n+1 $. Let $ E_v $ denote the event of $ \{s_{n+1},\cdots,s_{2n+1}\} = \{v_1,\cdots,v_{n+1} \}$. Then, under $E_v$, for $ i = 1,\cdots, k$, we have
$$\mathbb{P}(s_{2n+1}=\tilde{v}_i|E_v) = \frac{n_i}{n+1},$$
due to the exchangeability of $s_i$'s, 

We also note that under $ E_v$ and $ s_{2n+1} = \tilde{v}_i $ we have from equation (\ref{eq:p_val}),
\begin{equation}\label{eq: p_update}
 p(Z_{2n+1}) = \frac{\sum_{l=1}^i n_l}{n+1}. 
 \end{equation}
Then, for any $ t \in [0,1] $ and $ i = 1, \cdots, k $, we have
\begin{equation}
\mathbb{P}(p(Z_{2n+1}) \leq t|E_v,s_{2n+1} = \tilde{v}_i) =
\begin{cases}
0 & \text{if } t < \frac{\sum_{l=1}^i n_l}{n+1}, \\
1 & \text{o.w.} 
\end{cases}
\end{equation}
Thus, for any $ i = 1, \cdots, k$ and $\frac{\sum_{l=1}^{i-1} n_l}{n+1} \leq t < \frac{\sum_{l=1}^{i} n_l}{n+1}$, we have 
\begin{align*}
    & \qquad\mathbb{P}(p(Z_{2n+1}) \leq t|E_v) \\
    & = \sum_{l=1}^{k} \mathbb{P}(p(Z_{2n+1}) \leq t|E_v,s_{2n+1} = \tilde{v}_l) \cdot \mathbb{P}(s_{2n+1} = \tilde{v}_l|E_v) \\
    & = \frac{\sum_{l=1}^{i-1} n_l}{n+1} \leq t.
\end{align*}
By taking the expectation on the above inequality, it follows that the conformal $p$-value $ p(Z_{2n+1})$ is marginally valid. 

Specifically, if conformity scores $\{s_i\}_{i=n+1}^{2n+1}$ are distinct surely,  then $k = n+1$ and $n_i = 1$ for $i=1, \ldots, n+1$. Thus, 
\begin{center}
    $ \mathbb{P}\left(p(Z_{2n+1}) \leq t|E_v\right) = \frac{i-1}{n+1}, \textnormal{ if } \frac{i-1}{n+1} \leq t < \frac{i}{n+1}, $
\end{center}
that is, 
$$ p(Z_{2n+1})|E_v \sim U\Bigl\{\frac{1}{n+1},\cdots,1\Bigl\}. $$
This completes the proof.
\end{proof}

\subsection{Proof of Proposition 2}

\begin{proof}
For any given $y \in \mathcal{Y}$, the corresponding (class-specific) conditional conformal $p$-value is given by
\begin{equation}\label{eq:conditionalpval_prop}
    p(X_{2n+1}|y) = \frac{1}{n_y + 1} \left[ \sum_{i\in \mathcal{I}_y}\mathbb{I} \{s_{i} \leq s_{2n+1}\} + 1\right],
\end{equation}
where $\mathcal{I}_{y} = \{i:(X_i,Y_i) \in \mathcal{D}_{cal}, Y_i = y\}$, $ n_y = |\mathcal{I}_{y}| $,
$ s_{i} = s(X_i,y) $ for $i \in \mathcal{I}_{y}$, and $s_{2n+1} = s(X_{2n+1},y) $.
Given $I_y $ and $Y_{2n+1}=y$, $(X_i)_{i \in \mathcal{I}_y} \cup \{X_{2n+1}\}$ are exchangeably distributed, which is due to the assumption that  $ (X_i,Y_i)_{i=1}^{2n+1}$ are exchangeably distributed. Using the similar arguments as in the proof of Proposition 1, for any given values of $v_1, \cdots, v_{n_y + 1} $, suppose that they can be arranged as  $\tilde{v}_1 < \cdots < \tilde{v}_l$
with repetitions $ m_i $ of $ \tilde{v}_i $ such that $ \sum_{i=1}^{l} m_i = n_y + 1 $. 

Let $ E_v $ denote the event $ \{s_i\}_{i \in \mathcal{I}_{y}} \cup \{s_{2n+1}\} = \{ v_1,\cdots,v_{n_y + 1}\} $. Then, given $E_v, I_y $, and $ Y_{2n+1} = y $,  we have 
$$ \mathbb{P}(s_{2n+1} = \tilde{v}_i|E_v, I_y,Y_{2n+1} = y) = \frac{m_i}{n_y + 1}$$
for  $i = 1,\cdots, l$ and $y = 1,\cdots, K$, due to exchangeability of $ s_{i} $, $ i \in I_y \cup \{2n+1\} $ given $I_y$,  which in turn is due to exchangeability of $(X_i, Y_i)_{i=1}^{2n+1}$. Note that given $ E_v $,  $ I_y $,  $ Y_{2n+1} = y $, and $s_{2n+1} = \tilde{v}_i$, we have from equation (\ref{eq:conditionalpval_prop}),

$$ p(X_{2n+1}|y) = \frac{\sum_{j=1}^{i} m_j}{n_y+1}.$$
Thus, for any $ t \in [0,1] $ and $ i = 1,\cdots,l $,
\begin{equation}
\mathbb{P}\left(p(X_{2n+1}|y) \leq t \big| E_v, I_y, Y_{2n+1} = y, s_{2n+1} = \tilde{v}_i \right) =
\begin{cases}
0 & \text{if } t < \frac{\sum_{j=1}^{i} m_j}{n_y+1}, \\
1 & \text{o.w.} 
\end{cases}
\end{equation}
Then, for any given $ i = 1,\cdots,l $ and $\frac{\sum_{j=1}^{i} m_j}{n_y+1} \leq t < \frac{\sum_{j=1}^{i+1} m_j}{n_y+1}$, we have 
\begin{align*}
    & \qquad \mathbb{P}\left (p(X_{2n+1}|y) \leq t \big | E_v,I_y,Y_{2n+1} =y \right) \\
    & = \sum_{j=1}^{l} \mathbb{P}\left(p(X_{2n+1}|y) \leq t \big |E_v,I_y,Y_{2n+1}=y,s_{2n+1} = \tilde{v}_i\right ) \cdot \mathbb{P} \left(s_{2n+1} = \tilde{v}_i \big|E_v,I_y,Y_{2n+1}=y \right) \\
    & = \frac{\sum_{j=1}^{i} m_j}{n_y+1} \leq t.
\end{align*}
By taking expectation, it follows that $ p(X_{2n+1}|y) $ is conditionally valid given $ Y_{2n+1 } = y$.
\end{proof}

\subsection{Proof of Proposition 3}

\begin{proof}
Consider Procedure 1-3 based on marginal conformal $p$-values. Note that among the tested hypotheses $H_1, \ldots, H_K$, there is exactly one hypothesis $ H_{Y_{2n+1}}$ to be true. Thus, the FWER of Procedure 1-3 are all equal to
$$ \mathbb{P}(\text{reject } H_{Y_{2n+1}}) \leq \mathbb{P}\left(p(X_{2n+1},Y_{2n+1}) \leq \alpha\right) \leq \alpha,$$
where the last inequality follows by Proposition 1. 

\noindent Specifically,  for Procedure 3, if the conformity scores $ \{s_i\}_{i=n+1}^{2n+1} $ are distinct surely, by Proposition 1, we have 
\begin{align*}
    & \qquad \text{FWER}  = \mathbb{P}(\text{reject } H_{Y_{2n+1}}) 
         = \mathbb{P}\Bigl\{p(X_{2n+1},Y_{2n+1}) \leq \alpha \Bigl\} ~\geq \alpha - \frac{1}{n+1},
\end{align*}
the desired result.
\end{proof}

\subsection{Proof of Theorem 1}

\begin{proof}
Note that the prediction set derived from Algorithm 1 is given by
$C(X_{2n+1}) = A_1 \cap A_2.$
Thus, by Proposition 1,  $$ \mathbb{P}(Y_{2n+1} \in C(X_{2n+1})) \geq \mathbb{P}(p(X_{2n+1}),Y_{2n+1}) > \alpha) \geq 1 - \alpha.$$

Similarly, for Algorithm 2, its prediction set is given by $C(X_{2n+1}) = \{ y \in \mathcal{Y}: p(X_{2n+1}, y) > \alpha\}.$
By Proposition 1, it is easy to check that 
\begin{align*}
    \mathbb{P}(Y_{2n+1} \in C(X_{2n+1}))  = \mathbb{P}(p(X_{2n+1},Y_{2n+1}) > \alpha) \geq 1 - \alpha. 
\end{align*}
Specifically, if the conformity scores $\{s_i\}_{i=1}^{2n+1}$ are distinct surely, for Algorithm 2, we have
$$ \mathbb{P}(Y_{2n+1} \in C(X_{2n+1})) = 1 - \mathbb{P}(p(X_{2n+1},Y_{2n+1}) \leq \alpha) \leq 1-\alpha + \frac{1}{n+1}.$$
This completes the proof.
\end{proof}

\subsection{Proof of Proposition 4}
\begin{proof}
Consider Procedure 1-3 based on conditional conformal $ p$-values. For any $ y = 1,\cdots, K $, given $ Y_{2n+1} = y $,  the conditional FWER of Procedure 1-3 are all equal to
$$ \mathbb{P}(\text{reject } H_y|Y_{2n+1} = y) \leq \mathbb{P}(p(X_{2n+1}|y) \leq \alpha|Y_{2n+1} = y) \leq \alpha,$$
where the inequalities follow the definitions of Procedure 1-3 and Proposition 2. 

\noindent Specifically, for Procedure 3,  if the conformity scores $ \{s_{i}\}_{i\in \mathcal{I}_y \cup \{2n+1\}} $ are distinct surely, then by Proposition 2, 
the FWER conditional on $I_y$ and $Y_{2n+1} = y$ is equal to
\begin{align*}
    \mathbb{P}\Bigl\{ \text{reject}~ H_y ~ \big | I_y, Y_{2n+1} = y \Bigl\}  = \mathbb{P}(p(X_{2n+1}|y) \leq \alpha|I_y, Y_{2n+1} = y) \geq \alpha - \frac{1}{n_y + 1}.
\end{align*}
This completes the proof.
\end{proof}

\subsection{Proof of Theorem 2}

\begin{proof}
By using Proposition 4 and the similar arguments as in the proof of Theorem 1, the prediction sets $C(X_{2n+1}|y)$ derived from Algorithm 1 and 2 based on the conditional conformal $p$-values $p(X_{2n+1}|y)$ all satisfy,
$$\mathbb{P}\Bigl\{ Y_{2n+1} \in C(X_{2n+1} |y)~ \big | I_y, Y_{2n+1} = y \Bigl\} ~\geq 1 - \alpha$$
for any $y = 1, \ldots, K $.
\noindent Specifically, if the conformity scores $ \{s_{i}\}_{i\in \mathcal{I}_y \cup \{2n+1\}} $ are distinct surely, for Algorithm 2, we have
\begin{align*}
    \mathbb{P}\Bigl\{ Y_{2n+1} \in C(X_{2n+1} |y)~ \big | I_y, Y_{2n+1} = y \Bigl\} & = \mathbb{P}\Bigl\{ p(X_{2n+1}|y) > \alpha ~\big | I_y, Y_{2n+1} = y\Bigl\} \\
    & \leq 1 - \alpha + \frac{1}{n_y + 1},
\end{align*}
the desired result.
\end{proof}

\end{document}